\documentclass[12pt]{article}
\usepackage{epsfig}
\begin{document}
\newcommand{\ds}{\displaystyle}

\begin{center}
{\large \bf $\eta - \eta'$ in a hot and dense medium}
\end{center}

\begin{center}
P. Costa\footnote{pcosta@teor.fis.uc.pt}, M. C. Ruivo\footnote{maria@teor.fis.uc.pt}

{\em  Departamento de F\'{\i}sica
da Universidade de Coimbra

P - 3004 - 516 Coimbra, Portugal}

\vspace{0.3cm}

Yu. L. Kalinovsky\footnote{kalinov@nusun.jinr.ru}

{\em  Laboratory of Information Technologies,

Joint Institute for Nuclear Research, Dubna, Russia}
\end{center}
\vspace{3cm}

\begin{abstract}
The behavior of $\eta$ and $\eta'$ in hot strange quark matter in weak equilibrium 
with temperature, is investigated within the SU(3) Nambu-Jona-Lasinio [NJL] model.
Possible manifestations of restoration of symmetries, by temperature or density, 
in the behavior of $\eta$ and $\eta'$  are discussed. The role played by the combined 
effect of temperature and density in the nature of the phase transition and meson 
behavior is also analyzed.
\end{abstract}

\newpage
\section{Introduction}
\hspace*{\parindent}Understanding the  QCD phase diagram is one of the great challenges in the physics of strong interactions.
  It is expected that under extreme conditions (high density and or temperature)  phase transitions occur
leading to a plasma of deconfined quarks and gluons [QGP] \cite{kanaya,rhic,cern}.
  While  the phase transition at zero chemical potential and finite temperature is accepted to be second order or crossover, there  are indications that the
phase transition with finite chemical potential and zero temperature is first order.  Experimental and theoretical efforts have been done  in order to explore the $\mu-T$ phase boundary. Recent Lattice results indicate a critical "endpoint", connecting the  first order phase transition with the crossover region at $T_E=160\pm35\mbox{MeV}\,,\mu_E=725\pm35\mbox{MeV}$
\cite{Fodor}.
Besides restoration of chiral symmetry and deconfinement, the rich content of the QCD phase diagram
has been recently explored in other directions. For instance, in high density matter, for which neutron stars
 provide an excellent laboratory, a color locked phase with  kaon condensates is expected to occur.

Much attention has also been paid to the  question of which symmetries are restored. In the limit of vanishing
 quark masses the QCD Lagrangian has 8 Goldstone bosons, associated with the dynamical breaking of chiral symmetry.
  In order to give a finite mass to the mesons chiral symmetry is broken {\it ab initio} by giving current masses
 to the quarks, therefore in the high density/temperature regime an indication of the restoration of chiral symmetry
 would be that the mesons became again almost Goldstone bosons. However, the mass of $\eta'$ has a different origin than the masses of the other pseudoscalar mesons and it cannot be regarded as the remanent of a Goldstone boson.
The problem of   the non-existence of the ninth Goldstone boson, predicted by the quark model, was solved by assuming that the QCD Lagragian has a $U_A (1)$ anomaly. The mass of the $\eta' $ is obtained by explicitly breaking the $U_A (1)$ symmetry, for instance by instantons. 

 The study of the behavior of mesons  in hot and dense matter is  an important issue since they might provide a
signature of the phase transition and give indications about which symmetries are restored.  There is a long standing question  \cite{shuryak} about whether only SU(3) chiral symmetry is restored or both SU(3) and $U_A(1)$
symmetries are restored. The behavior of $\eta'$ in medium or of related observables like the
topological susceptibility might help to decide between these scenarios.  A decrease of the $\eta'$ mass
in medium could manifest in the increase of the $\eta'$ production cross section, as compared to that for
$pp$ collisions \cite{kapusta}.

This paper is devoted to investigate the in medium behavior of the $\eta \mbox{ and } \eta'$ mesons in hot
and dense matter in the framework of the SU(3) Nambu-Jona-Lasinio [NJL] \cite{njl} model including the 't Hooft
interaction which breaks the $U_A(1)$ symmetry. At variance with the other pseudoscalar SU(3) mesons there is not
much information about the behavior of these mesons in medium. Concerning the $\eta$ meson, the recent
discovery of mesic atoms might provide useful information in this concern \cite{sokol}.

The NJL model is an effective quark model where the gluonic degrees of freedom are supposed to be frozen and,
 besides its simplicity, has the advantage of incorporating important symmetries of QCD, namely chiral symmetry.
Since the model has no confining mechanism,  several drawbacks are well known. In what concerns our present problem,
 it should be noticed that the $\eta'$ mass lies above the $\bar q q$ threshold  and the model describes
this meson as $\bar q q$ resonance, which would have the unphysical decay in $\bar q q$ pairs. For this reason,  it has been proposed that in order to investigate
the possible restoration of $U_A(1)$ symmetry, instead of $m_{\eta'}$, the topological susceptibility $\chi$,
a more reliable quantity in this model, should be used \cite{ohta1}.

The behavior of SU(3) pseudoscalar mesons, in NJL models,  with temperature has been studied in \cite{njlT,bla}.
Different studies have been devoted to the behavior of pions and kaons at finite density in flavor symmetric \cite{maria} or asymmetric matter \cite{RuivoSousa,SousaRuivo,costa} . The combined effect of temperature and density  on kaons in quark matter simulating symmetric nuclear matter has been investigated in \cite{RSP}. 

In this work we investigate the phase transition and the behavior of $\eta, \eta'$ in strange quark matter in weak equilibrium at zero and finite temperature.   We present the model and formalism in the vacuum   (Section 2) and   at finite density and temperature (Section 3). In Section 4 we discuss present our results for the phase transition in the $T-\rho$ plane and in Section 5 we present our results and conclusions for the $\eta\,,\eta'$ behavior.


\section{Model and Formalism}

\hspace*{\parindent}We will consider in this paper the $U(1)_A$ anomaly in
the three  flavor Nambu - Jona - Lasinio model. The Lagrangian
we employ here can be used as an effective theory of QCD and
has the following form
\begin{eqnarray}
{  L} &=& \bar{q} \left( i \partial \cdot \gamma - \hat{m} \right) q
+ \frac{g_S}{2} \sum_{a=0}^{8}
\Bigl[ \left( \bar{q} \lambda^a q \right)^2+
\left( \bar{q} (i \gamma_5)\lambda^a q \right)^2
 \Bigr] \nonumber \\
&- &
\frac{g_V}{2} \sum_{a=0}^{8} \Bigl[
\left( \bar{q} \gamma_\mu \lambda^a q\right)^2+
\left( \bar{q}  (\gamma_\mu \gamma_5 )\lambda^a q\right)^2
 \Bigr]
 \nonumber \\
&+& g_D \Bigl[ \mbox{det}\bigl[ \bar{q} (1-\gamma_5) q \bigr]
  +  \mbox{det}\bigl[ \bar{q} (1-\gamma_5) q \bigr]\Bigr] \, .
\label{lagr}
\end{eqnarray}
Here $q = (u,d,s)$ is the quark field with three flavors, $N_f=3$, and
three colors, $N_c=3$. $\lambda^a$ are the Gell - Mann matrices, a=$0,1,\ldots , 8$,
${ \lambda^0=\sqrt{\frac{2}{3}} \, {\bf I}}$. The explicit symmetry breaking part (\ref{lagr})
contains the current quark masses $\hat{m}=\mbox{diag}(m_u,m_d,m_s)$. Note, that
$m_{u,d} \sim 10$ MeV, $m_s \sim$ 100 MeV at the scale of QCD ($\sim$ 1 GeV).
The last term in (\ref{lagr}) is the lowest dimensional operator and that
 has the $SU_L(3)\otimes SU_R(3)$ invariance but breaks the
$U_A(1)$ symmetry. This term is a reflection of the axial anomaly in QCD.
The model is fixed by the coupling constants $g_S, g_V, g_D$, the cutoff parameter $\Lambda$,
which regularizes momentum space integrals and current quark masses. In order to fix parameters we
will use the values of pseudoscalar meson masses, decay constants and the quark condensates
at zero temperature and baryon number density.

Using the definition of the determinant we can rewrite the Lagrangian (\ref{lagr}) as
\begin{eqnarray}
&& {  L} = \bar{q} \left( i \partial \cdot \gamma - \hat{m} \right) q
+
\frac{g_S}{2} \sum_{a=0}^{8}
\Bigl[ \left( \bar{q} \lambda^a q \right)^2+
\left( \bar{q} (i \gamma_5)\lambda^a q \right)^2
 \Bigr]
 \nonumber \\ & -&
\frac{g_V}{2} \sum_{a=0}^{8} \Bigl[
\left( \bar{q} \gamma_\mu \lambda^a q\right)^2+
\left( \bar{q}  (\gamma_\mu \gamma_5 )\lambda^a q\right)^2
 \Bigr]
 \nonumber \\
&+& \frac{g_D}{2} \Bigl[
\frac{1}{3} D_{abc} (\bar{q} \lambda^a q)(\bar{q} \lambda^b q)(\bar{q} \lambda^c q)
-D_{abc} (\bar{q} \lambda^a q)(\bar{q} (i\gamma_5) \lambda^b q)(\bar{q} (i\gamma_5)\lambda^c q)
\Bigr]
\end{eqnarray}
with summation over $a,b,c =(0,1,2,\ldots ,8)$. The structure constants $D_{abc}$
are coincident with SU(3) structure constants $d_{abc}$ and
$D_{0ab}=-\frac{1}{\sqrt{6}}\delta_{ab}$ and $D_{000}=\sqrt{\frac{2}{3}}$.
 Making the shift
\begin{eqnarray}
(\bar{q} \lambda^a q) \longrightarrow (\bar{q} \lambda^a q) + <\bar{q} \lambda^a q> \, ,
\end{eqnarray}
where $<\bar{q} \lambda^a q>$ is the vacuum expectation value, and keeping second order
terms of $(\bar{q} \lambda^a q)$, we get for pseudoscalar and scalar interactions (for simplicity,
we omit vector and axial - vector parts)
\begin{eqnarray}\label{NJL}
  {  L} =
\bar{q} \left( i \partial \cdot \gamma - \hat{m} \right) q +
\frac{1}{2}
\Bigl\{
(\bar{q} \lambda^a q)S_{ab}(\bar{q} \lambda^b q)
+(\bar{q} (i\gamma_5) \lambda^a q) P_{ab} (\bar{q} (i\gamma_5) \lambda^b q)
\Bigr\} \, ,
\end{eqnarray}
in terms of the projectors
\begin{eqnarray}
  S_{ab} &=& g_S \delta_{ab} + g_D D_{abc}<\bar{q} \lambda^c q> \, , \label{sab}\\
  P_{ab} &=& g_S \delta_{ab} - g_D D_{abc}<\bar{q} \lambda^c q> \, . \label{pab}
\end{eqnarray}

The hadronization procedure is made via the functional integral
\begin{eqnarray}\label{fint}
  Z \sim \int dq d\bar{q} \,\,  \mbox{exp}\left\{{i W_{eff}[q,\bar{q}]}\right\} \,
\end{eqnarray}
with the quark effective action
\begin{eqnarray}
&& W_{eff}[q,\bar{q}] = \int dx {  L}(x)
=
\int dx \Bigl[
\bar{q} \left( i \partial \cdot \gamma - \hat{m} \right) q
\nonumber \\ &&
+
\frac{1}{2}
(\bar{q} \lambda^a q)S_{ab}(\bar{q} \lambda^b q)
+ \frac{1}{2}(\bar{q} (i\gamma_5) \lambda^a q) P_{ab} (\bar{q} (i\gamma_5) \lambda^b q)
\Bigr]\, .
\end{eqnarray}
After integration over quark fields in (\ref{fint}),  we obtain the effective meson action
\begin{eqnarray}\label{act}
  W_{eff}[\varphi,\sigma] &=&
  - \frac{1}{2} \left( \sigma^a S^{-1}_{ab}\sigma^b \right)
  - \frac{1}{2} \left( \varphi^a P^{-1}_{ab}\varphi^b \right)
  \nonumber \\ &&
  -i \mbox{Tr} \,  \mbox{ln} \Bigl[ i (\gamma_\mu \partial_\mu ) - \hat{m} + \sigma_a \lambda^a
  + (i \gamma_5 )(\varphi_a \lambda^a) \Bigr] \, .
\end{eqnarray}
Here the symbol $\mbox{Tr}$ means the summation over discrete indices and integration over
continuous variables. The fields $\sigma^a$ and $\varphi^a$ are scalar and
pseudoscalar meson nonets,  respectively.

The first variation of the action (\ref{act}) leads to the Dyson - Schwinger equation,
\begin{eqnarray}\label{gap}
  M_i = m_i - 2g_{S} <\bar{q_i}q_i> -2g_{D}<\bar{q_j}q_j><\bar{q_k}q_k>\, ,
\end{eqnarray}
with $i,j,k =u,d,s$ cyclic. $M_i$ are constituent quark masses.
The equation (\ref{gap}) is a system of three coupled equations and the flavor mixing
occurs through the coupling constant $g_D$. It is an effect of anomaly contribution.
The quark condensates in (\ref{gap}) are determined by
\begin{eqnarray}
 <\bar{q}_i  q_i> =  -i \mbox{Tr} \frac{1}{\hat{p}-M_i} = -i \mbox{Tr}\left[ S_i(p) \right].
\end{eqnarray}
Here $S_i(p)$  is the quark Green function.

To  consider the meson mass spectrum,  we expand the effective action
(\ref{act}) over meson fields.
Keeping the pseudoscalar mesons only,
we  write the second order effective action as
\begin{eqnarray}\label{act2}
  W_{eff}^{(2)}[\varphi] = -\frac{1}{2}\varphi^a \left[ P_{ab}^{-1} - \Pi_{ab} (P) \right] \varphi^b
\end{eqnarray}
with the polarization operator $\Pi_{ab}(P)$, which in the momentum space has the form
\begin{eqnarray}
\Pi_{ab} (P) = i N_c \int \frac{d^4p}{(2\pi)^4}\mbox{tr}_{D}\left[
S_i (p+\frac{P}{2}) (\lambda^a)_{ij} (i\gamma_5) S_j (p-\frac{P}{2})(\lambda^b)_{ji} (i\gamma_5)
\right] \, ,
\end{eqnarray}
where now $\mbox{tr}_{D}$ is the trace over Dirac matrices.
In the result, we have obtained  an unnormalized inverse meson propagator
${  D}^{-1}_{ab} = P_{ab}^{-1} - \Pi_{ab}(P)$ or
\begin{eqnarray}\label{prop}
  {  D}_{ab} = \frac{1}{P_{ab}^{-1} - \Pi_{ab}(P)}\,
\end{eqnarray}
which is the 3$\times$3 matrix in the flavor space. We will neglect $\pi^0 - \eta$ and
$\pi^0 - \eta'$ mixing terms because they are proportional to $<\bar{q}_uq_u> - <\bar{q}_dq_d>$. 
We will do the approximation of neglecting also  this contribution in the medium. 
In the basis of $\eta - \eta'$ mesons we may find $P_{ab}$ and $\Pi_{ab} (P)$
\begin{eqnarray}
  P_{ab} =
  \left(  \begin{array}{cc} P_{00} & P_{08} \\ P_{08 } &P_{88}  \end{array} \right)
  \longrightarrow
 P^{-1}_{ab} = \frac{1}{\Delta}
  \left(  \begin{array}{cc} P_{88} & -P_{08} \\ -P_{08 } &P_{00}  \end{array} \right)
\end{eqnarray}
with the determinant $\Delta  = P_{00}P_{88} -P_{08}^2 $.
Then the inverse meson propagator  (\ref{prop}) can be presented in the explicit form
\begin{eqnarray}
{  D}^{-1}_{ab} &=&
\frac{1}{\Delta}
\left(  \begin{array}{cc} P_{88}- \Delta \Pi_{00} & -P_{08}- \Delta \Pi_{08}
\\ -P_{08 } - \Delta \Pi_{08} &P_{00}- \Delta \Pi_{88}  \end{array} \right)
   \nonumber \\
&=&
\frac{1}{\Delta}
\left(  \begin{array}{cc} {  A} & {  B}
\\ {  B} & {  C}  \end{array} \right)
\equiv \frac{1}{\Delta}
{  O}^{-1}\left(  \begin{array}{cc} \underline{  A} & 0
\\ 0 & \underline{  C}  \end{array} \right){  O}
\end{eqnarray}
with the  orthogonal  transformation matrix ${  O}$.
Let us suppose that the matrix ${  O}$ has the form
\begin{eqnarray}\label{mo}
  {  O}=\left(  \begin{array}{cc} \cos \theta & \sin \theta\\ -
  \sin \theta& \cos\theta  \end{array} \right) \, .
\end{eqnarray}
The value of the angle $\theta$ can be fixed from the condition
\begin{eqnarray}\label{tan}
  \tan 2\theta = \frac{2 {  B}}{{  C}-{  A}} \,
\end{eqnarray}
which guarantees us a diagonal form of ${  D}^{-1}_{ab}$.
From  (\ref{mo}) and (\ref{tan}) we may find $\underline{  A}$ and $\underline{  C}$
and write  ${  D}_{ab}$ as
\begin{eqnarray}
  {  D}^{-1}_{ab}
  = \frac{1}{\Delta}
{  O}^{-1}\left(  \begin{array}{cc} \underline{  A} & 0
\\ 0 & \underline{  C}  \end{array} \right){  O}
= \frac{1}{2\Delta}
{  O}^{-1}\left(  \begin{array}{cc} D_\eta^{-1} & 0
\\ 0 & D_{\eta'}^{-1}  \end{array} \right){  O}
\end{eqnarray}
where
\begin{eqnarray}
D_\eta^{-1} = \left( {  A}+{  C}\right) -  \sqrt{({  C}-{  A})^2+4{  B}^2}
\end{eqnarray}
and
\begin{eqnarray}
D_{\eta'}^{-1} =\left( {  A}+{  C}\right) +  \sqrt{({  C}-{  A})^2+4{  B}^2} \, .
\end{eqnarray}
The masses of the $\eta$ and $\eta'$ meson can now be determined by the conditions
\begin{eqnarray}\label{mesgap1}
 D_\eta^{-1}(M_\eta,{\bf 0}) =0 \, ,
\end{eqnarray}
\begin{eqnarray}\label{mesgap2}
 D_{\eta'}^{-1}(M_{\eta'},{\bf 0}) =0 \, .
\end{eqnarray}

To calculate the coupling constants, we will express ${  D}_{ab}$ directly in the
following form
\begin{eqnarray}
{  D}_{ab} =
{\Delta}
\left(  \begin{array}{cc} {  A} & {  B}
\\ {  B} & {  C}  \end{array} \right)^{-1}
= \frac{\Delta}{D}\left(  \begin{array}{cc} {  C} & -{  B}
\\ -{  B} & {  A}  \end{array} \right) \, .
\end{eqnarray}
Here
\begin{eqnarray*}
 \Delta &=& P_{00}P_{88} -P_{08}^2 \, , \\
 D &=& {  A} {  C} -{  B}^2 \, \nonumber \\
&=& \Delta ( 1- \mbox{tr}(P \Pi) + \mbox{det} P \mbox{det} \Pi  ) \,
\end{eqnarray*}
and the effective action (\ref{act2}) now is
\begin{eqnarray}
  W_{eff}^{(2)}[\varphi] = -\frac{1}{2} \varphi^a { D}_{ab} \varphi^b
\end{eqnarray}
with
\begin{eqnarray}
{ D}_{ab}
= \frac{1}{\tt  D}\left(  \begin{array}{cc} {  C} & -{  B}
\\ -{  B} & {  A}  \end{array} \right) \, .
\end{eqnarray}
It has  singularities when ${\tt  D}$ equals zero.
Making the pole approximation, we  expand ${\tt  D}$ over $P^2$:
\begin{eqnarray}
  {\tt  D}(P^2) = {\tt  D}(P^2=M^2) +
  \frac{\partial {\tt  D}}{\partial P^2 }_{|_{P^2=M^2}} \, (P^2-M^2) \,.
\end{eqnarray}
The first term equals  zero when $P^2=M_\eta^2 $ (or $P^2=M_{\eta'}^2 $).
The second term   can be rewritten in the rest frame as
\begin{eqnarray}
  {\tt  D}(P^2) = \frac{1}{2M}
  \frac{\partial {\tt  D}}{\partial P_0 }_{|_{P_0=M}} \, (P^2-M^2) \,.
\end{eqnarray}
In the case $P^2=M_\eta^2$ we have the following expression for the meson propagator
\begin{eqnarray}
  {D}_{ab} = \frac{2M_\eta}{\left(\frac{\partial {\tt  D}}{\partial P_0 }
\right)_{|_{P_0=M_\eta}}}
  \, \frac{1}{P^2-M_\eta^2} \,
\left(  \begin{array}{cc} {  C} & -{  B}
\\ -{  B} & {  A}  \end{array} \right) \,.
\end{eqnarray}
The matrix
\begin{eqnarray}\label{M}
  M_{ab} = - g_{a\eta} \frac{1  }{P^2-M_\eta^2}g_{b\eta}
\end{eqnarray}
 allows to parameterize the meson propagator in terms of coupling constants
$g_{i\eta}$ with $i=0,8$.
Using (\ref{M}),  we obtain
\begin{eqnarray}
g^2_{0\eta} &=& -\frac{2M_\eta}{\left(\frac{\partial {  D}}{\partial
P_0 } \right)_{|_{P_0=M_\eta}}} {  C}
\, , \\
g^2_{8\eta} &=& -\frac{2M_\eta}{\left(\frac{\partial {  D}}{\partial
P_0 } \right)_{|_{P_0=M_\eta}}} {  A}
\, , \\
g_{0\eta}g_{8\eta} &=& \frac{2M_\eta}{\left(\frac{\partial {
D}}{\partial P_0 } \right)_{|_{P_0=M_\eta}}}
{  B} \, .
\end{eqnarray}
From these coupling constants we may calculate
\begin{eqnarray}
  g_{\eta q\bar{q}} &=& \sqrt{\frac{2}{3}} g_{0\eta} +\frac{1}{\sqrt{3}}
g_{8\eta} \, ,  \\
g_{\eta s\bar{s}} &=& \sqrt{\frac{2}{3}} g_{0\eta} -\frac{2}{\sqrt{3}}
g_{8\eta} \, .
\end{eqnarray}
The coupling constants $g_{0\eta'},g_{8\eta'},g_{\eta'q\bar{q}
},g_{\eta's\bar{s} }$ can be evaluated
in the same way by the replacement $M_\eta \rightarrow M_\eta'$.

Now we consider the properties of $\eta$ and $\eta'$ mesons.

First we solve the gap equation (\ref{gap}). Then
the matrices ${  A, B, C, D}$ depend  on the solutions of $M_i$ and
contain the integrals:
\begin{eqnarray}\label{i1}
I_1^i &=& i N_c \int \frac{d^4p}{(2\pi)^4} \, \frac{1}{p^2-M_i^2}
\end{eqnarray}
and
\begin{eqnarray}\label{i2}
I_2^{ii}(P) &=& i N_c \int \frac{d^4p}{(2\pi)^4} \, \frac{1}{(p^2-M_i^2)((p+P)^2-M_i^2)},
\end{eqnarray}
which are divergent. To calculate the physical observables 
we introduce the cut - off parameter $\Lambda$. Putting
the integrals to (\ref{mesgap1}) we may
calculate $M_\eta$. But,
since NJL does not confine the quarks and the $\eta'$ is above
the $\bar{q}_u q_u$ and $\bar{q}_d q_d$ continuum,  the meson propagator (\ref{mesgap2})
 has complex poles.
We assume (see  details in \cite{bla} )
that the equation for the poles in the propagator has solutions of the form:
\begin{equation}
P_0 = M_{\eta'} - \frac{1}{2} i \Gamma,
\end{equation}
\noindent where  $M_{\eta'}$ is the mass and $\Gamma$ is the width of the $\eta'$ - resonance.

In order to avoid the complexity of the  analytical continuation, we calculate
the integrals (\ref{i2}) in an approximate way, which means that
we neglect  $\Gamma$  that occurs in the denominator of $I_2^{ii} (P_0)$
and consider the small value of the width. Since this last integral is complex,
a straightforward calculation leads to the replacement in ${  A, B, C, D}$:
\begin{equation}
P_0^{2}I_2^{ii}(P_0) \longrightarrow \left[
P_0^{2} \mbox{Re} I_2^{ii}(P_0) + P_0%
\Gamma \mbox{Im} I_2^{ii}(P_0) \right]
+ i \left[ P_0^{2} \mbox{Im} I_2^{ii}(P_0) - P_0 \Gamma
\mbox{Re} I_2^{ii}(P_0)\right] \vert_{P_0^{2}=M_{\eta'}}.
\end{equation}

Therefore from the zeros of the real and imaginary parts of the propagator we get the mass
and the width of $\eta'$.

For our numerical calculations we use the parameter set:

\noindent
$m_u = m_d =        5.5$ MeV,
$m_s =            140.7$ MeV,
$g_S \Lambda^2 =   3.67$,
$g_D \Lambda^5 = -12.36$ and
$\Lambda =        602.3$ MeV

\noindent
that has been determined by fixing the conditions:

\noindent
$M_\pi = 135.0$ MeV,
$M_K   = 497.7$ MeV,
$f_\pi =  92.4$ MeV and
$M_{\eta'}= 960.8$ MeV.

\noindent
 Here $f_\pi$ is the pion decay constant \cite{njlT,bla}.

Also, we have:

\noindent
$M_{\eta}= 514.8$ MeV, $\theta (M_{\eta}^2) = -5.8^{\circ}$, $ g_{\eta \bar{u}u} =   2.29$,
$g_{\eta \bar{s}s} = -3.71$

\noindent
and

\noindent
$M_{\eta'}= 960.8$ MeV, $\theta (M_{\eta'}^2) = -43.6^{\circ}$, $ g_{\eta'\bar{u}u} =  13.4$,
$g_{\eta' \bar{s}s} = -6.72$.

Note that the $\eta'$ - meson always lies above the
quark - antiquark threshold. Since this is an artifact due to the lack of confinement
it is a resonant state. Nevertheless, we use  $M_{\eta'}$ as an input parameter,
because as mentioned in \cite{ohta1},
the chiral susceptibility is related with $M_{\eta'}$ by Witten - Veneziano mass formula
\cite{witten,veneziano}
\begin{eqnarray}\label{sus}
\frac{2 N_f}{f_\pi^2} \chi = M_{\eta}^2+ M_{\eta'}^2-2 M_{K}^2
\end{eqnarray}
and takes the value of $M_{\eta'}$ which is closed to the experimental value.
In this formula $N_f=3$  and the chiral susceptibility equals $\chi = (177.2 \mbox{MeV})^4 $.
The lattice calculations  give the number $\chi = (175 \pm 5 \mbox{MeV})^4 $.

For the quark condensates we have
$<\bar{q}_u q_u> = <\bar{q}_d q_d> = - (241.9 \mbox{MeV})^3$ and
$<\bar{q}_s q_s> = - (257.7 \mbox{MeV})^3$.


\section{The finite temperature and density case}

\hspace*{\parindent}The generalization of the NJL model to the  finite temperature and density case
can be done if we introduce the thermal Green function, which for a system of quarks $q_i$ at
given temperature $T$ and chemical potential $\mu_i$ is

\begin{eqnarray}\label{termal}
S_i (\vec x -\vec x^{\prime},\tau -\tau^{\prime})& = &\frac{i}{\beta}\sum_{n}
e^{-i \omega_n (\tau - \tau^{\prime})}
\int\frac{d^3p}{(2 \pi)^3} \frac{e^{-i\,\vec p\,(\vec x -\vec x%
^{\prime})}}{\gamma_0 (i \omega_n + { \mu_i}) - \vec \gamma . \vec p
- M_i}\,,
\end{eqnarray}
where $\beta = 1/T$, $T$ is the temperature and the sum is made over the Matsubara frequencies
$\omega_n=(2n+1)\pi T$, $n=0,\pm 1, \pm 2, \ldots$,
so that $p_0 \longrightarrow i\omega_n + \mu_i$ with a chemical potential $\mu_i$.
Instead of integration over $p_0$ we have now the sum
over Matsubara frequencies.
$E_{i}=\sqrt{{\tt p}^2+M_{i}^2}$ is the quark energy.

 Because in the gap equation (\ref{gap})
instead of $S_i(p)$ we put the thermal Green functions (\ref{termal}),
the current quark masses $M_i$ will now
depend on the temperature and chemical potential
\begin{equation}
{M_i\,=\,m_i\,-2\,g_S\,<<\bar q_i\, q_i>>\,-\,2\,g_D\,<<\bar q_j\, q_j>>
<<\bar q_k\, q_k>>}
\end{equation}
where $<<\bar q_i\, q_i>> $ are the quark condensates at finite $T$ and
chemical potential $\mu_i$. The condensates
are expressed in the terms of the integral $I_1^i(T,\mu_i)$ which is calculated
by the substitution of (\ref{termal}) in (\ref{i1})
\begin{eqnarray}\label{firstt}
I_1^i (T, \mu_i) = - \frac{N_c}{4\pi^2} \int \frac{{\tt p}^2 d{\tt p}}{E_i} \left(
n^+_i - n^-_i
\right),
\end{eqnarray}
with the Fermi distribution functions
\begin{eqnarray}
n_i^\mp = \Bigl[ 1 + \mbox{exp} \left[ \beta (E_i \pm \mu_i)\right] \Bigr]^{-1}
\end{eqnarray}
The integral $I_2^{ii}(P_0, T, \mu_i)$ may be found in the same fashion
\begin{eqnarray}\label{sint}
I_2^{ii}(P_0, T, \mu_i) =
&& - \frac{N_c}{2\pi^2} {\mathcal{P}}\int \frac{{\tt p}^2 d {\tt p}}{E_i} \,\,
\frac{1}{P_0^2-4 E_i^2} \left( n^+_i - n^-_i\right)
\nonumber \\
&& - i  \frac{N_c}{4\pi} \sqrt{ 1- \frac{4 M_i^2}{P_0^2}  }
\left(n^+_i(\frac{P_0}{2}) - n^-_i (\frac{P_0}{2})\right) \,.
\end{eqnarray}

Using these integrals we  solve equations (\ref{mesgap1}) and (\ref{mesgap2})
at given temperature and chemical potential taking into account also
the relations
\begin{eqnarray}
P_{00}(T,\mu_i)&=&g_{S}-\frac{2}{3}g_{D}\left( <<\bar{q}_{u}\,q_{u}>>+<<\bar{q}%
_{d}\,q_{d}>>+<<\bar{q}_{s}\,q_{s}>>\right) , \\
P_{88}(T,\mu_i)&=&g_{S}+\frac{1}{3}g_{D}\left( 2<<\bar{q}_{u}\,q_{u}>>+2<<\bar{q}%
_{d}\,q_{d}>>-<<\bar{q}_{s}\,q_{s}>>\right) ,\\
P_{08}(T,\mu_i)&=&P_{80}=\frac{1}{3\sqrt{2}}g_{D}\left( <<\bar{q}_{u}\,q_{u}>>+<<\bar{q}%
_{d}\,q_{d}>>-2<<\bar{q}_{s}\,q_{s}>>\right) ,
\end{eqnarray}
and
\begin{eqnarray}
\Pi_{00}(P_0,T,\mu_i)&=&\frac{2}{3}\left[ J_{uu}(P_0,T,\mu_i)+J_{dd}(P_0,T,\mu_i)+J_{ss}(P_0,T,\mu_i)\right] , \\
\Pi_{88}(P_0,T,\mu_i)&=&\frac{1}{3}\left[J_{uu}(P_0,T,\mu_i)+J_{dd}(P_0,T,\mu_i)+4J_{ss}(P_0,T,\mu_i)\right],\\
\Pi_{08}(P_0,T,\mu_i)&=& \frac{\sqrt{2}}{3}\left[J_{uu}(P_0,T,\mu_i)+J_{dd}(P_0,T,\mu_i)
-2J_{ss}(P_0,T,\mu_i)\right],
\end{eqnarray}
where 
\begin{equation}
J_{ii}(P_0,T,\mu_i)=4(I_1^i+\frac{P_0^2}{2}I_2^{ii}).
\end{equation}


\section{Phase transition}
\hspace*{\parindent}To consider the density effects,
in accordance with \cite{RuivoSousa,SousaRuivo,RSP,Hiller,Ruivo,costaruivo},
we  concentrate here on the case
of asymmetric quark matter with strange quarks in chemical
equilibrium maintained by weak interactions and with charge neutrality,  by imposing the following
constraints on the chemical potentials of quarks and electrons and on its  densities:

\begin{equation}\label{constr1}
\mu_{d}=\mu_{s}=\mu_{u}+\mu_{e}\,\, \mbox{ and }\,\,\,\frac{2}{3}\rho_{u}-\frac{1}{3}(\rho_{d}+\rho_{s})-\rho_{e}=0,
\end{equation}
with
\begin{equation}\label{constr2}
\rho_{i}=\frac{1}{\pi^{2}}(\mu_{i}^{2}-M_{i}^{2})^{3/2}\theta(%
\mu_{i}^{2}-M_{i}^{2})\,\,\mbox{ and }\,\,\,\rho_{e}=\frac{\mu_{e}^{3}}{3\pi^{2}}.
\end{equation}

As discussed by several authors, this  version of the NJL model exhibits,
at $T = 0$ MeV a first order phase transition \cite{costa,RSP,buballa}.

For the numerical calculations and discussions of the nature of the phase transition
we use the expression for the pressure at given
temperature which is connected with the thermodynamical potential as
\begin{equation}\label{pressure}
P(\rho, T) = - \left[ \Omega(\rho, T) - \Omega(0, T) \right],
\end{equation}
where the thermodynamic potential  has the following form
\begin{equation}\label{tpot}
\Omega (\rho ,T)= {\mathcal{E}} - TS - \sum_{i=u,d,s} \mu _{i} N_{i} \,
\end{equation}
and contains  the internal energy $\mathcal{E}$, the entropy $S$
and the number of the $i$th quark $N_i$.


\begin{figure}
\begin{center}
  \begin{tabular}{cc}
    \hspace*{-0.5cm}\epsfig{file=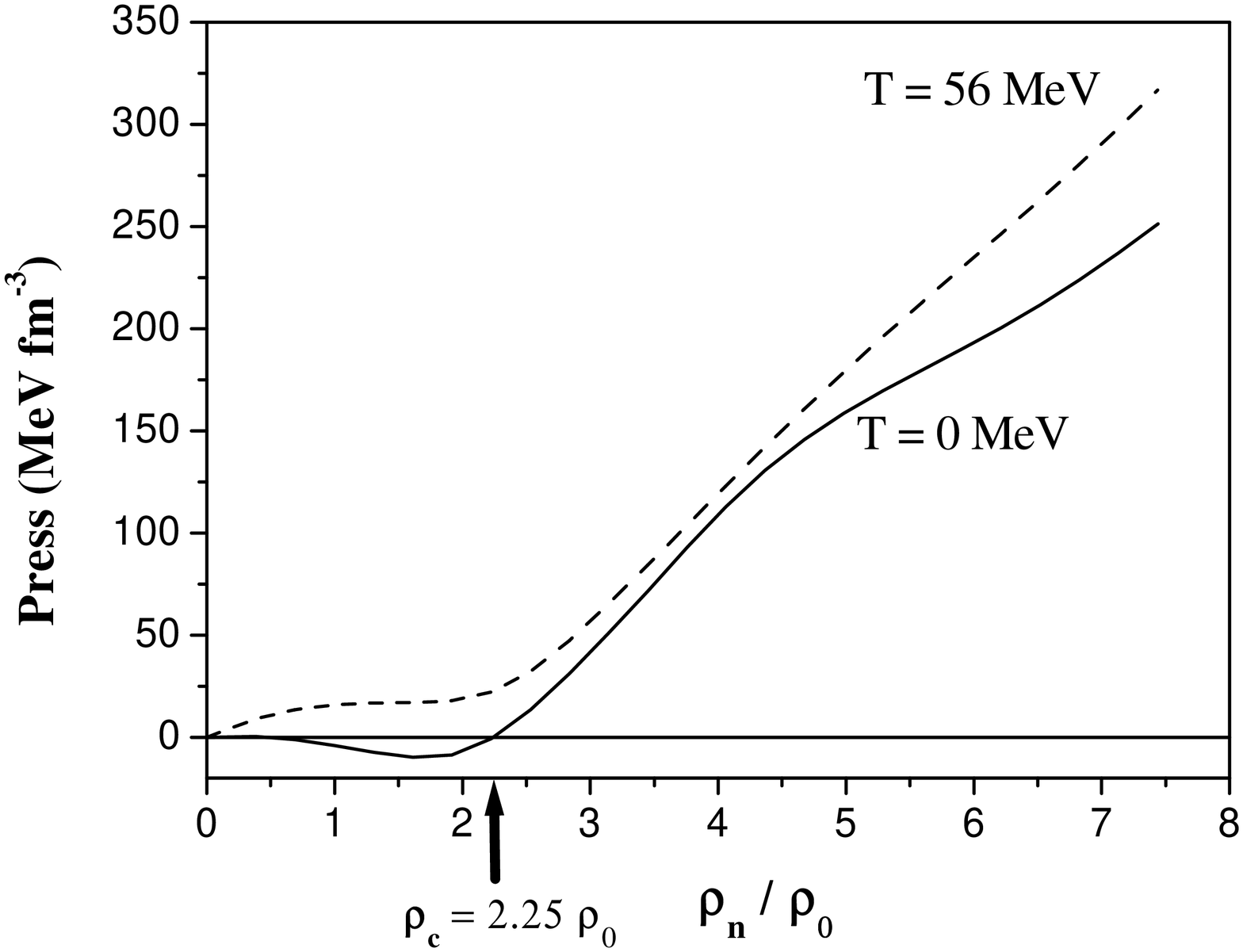,width=9cm,height=7cm} &
    \epsfig{file=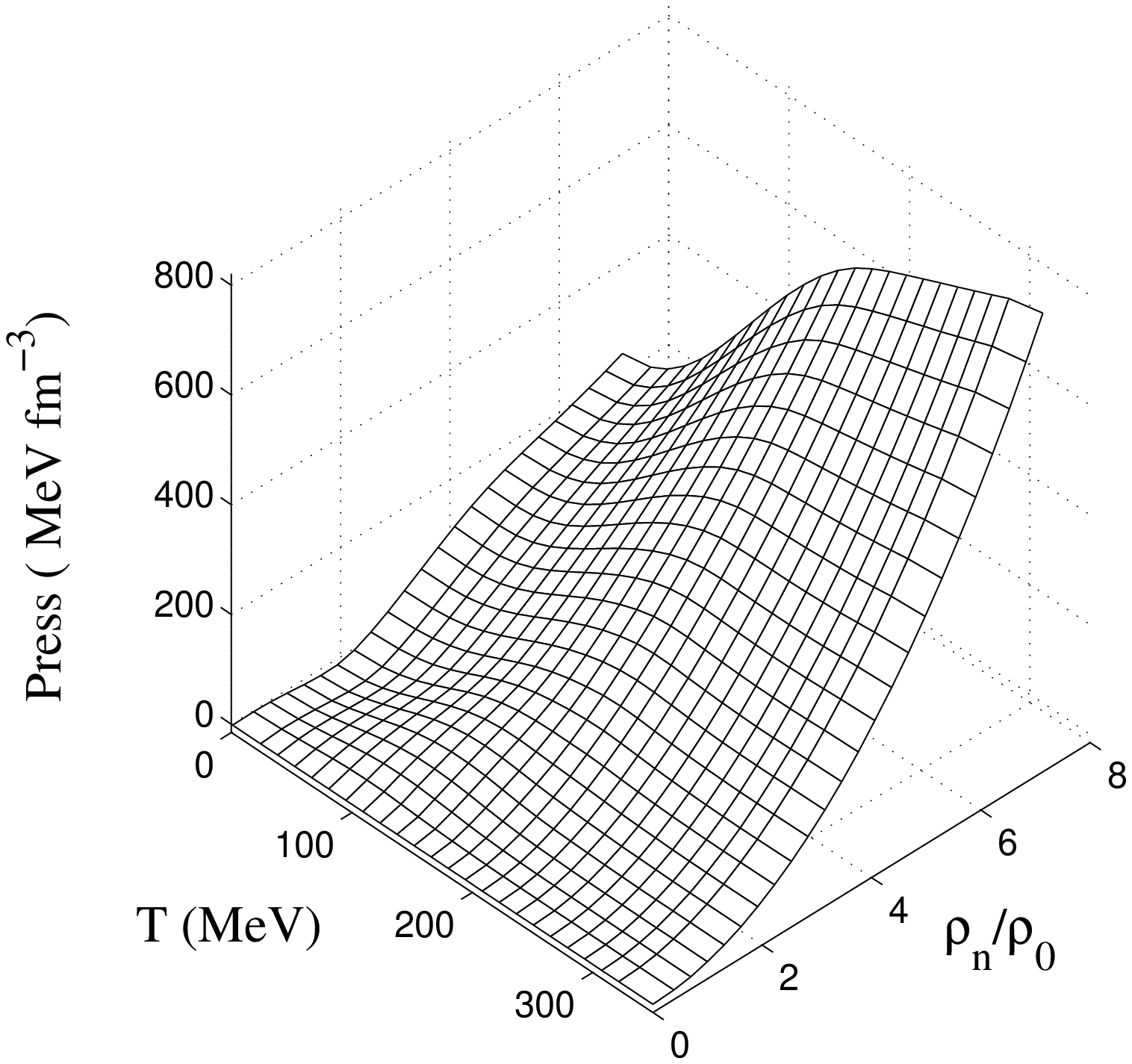,width=9cm,height=7cm} \\
   \end{tabular}
\end{center}
\caption{Left panel: Pressure $ T = 0$ MeV (1st order phase transition) and $ T = 56$ MeV
(2nd order phase transition);
Right panel: Pressure as function of the density and temperature.
At $ T < 56$ MeV we have a first order phase transition.
After that, the phase transition becomes of second order.}
\end{figure}

As a matter of fact, the energy per particle of the quark system has two minimum, corresponding
to the zeros of the pressure, the minimum at  $\rho_n = 2.25 \rho_0$
being an absolute minimum.
This implies that the phase transition  in this model is a first
order one. The results are presented in Figure 1 (left panel) where we have plotted
the pressure in terms of $\rho_n/\rho_0$, where $\rho_n$ is the neutron
matter density and $\rho_0 = 0.17$ fm$^{-3}$ is the normal nuclear density.

At finite temperature we conclude that for $T<56$ MeV we still have
a first order phase transition,
but for $T > T_{cl}= 56$ MeV the phase transition is of second order.
 
Below this temperature there is a range of densities ($\rho_n < 2.25 \rho_0$) where
the pressure and/or the compressibility are negative meaning that the system is in a mixing phase:
a low density ($\rho_n^l$) phase  with massive quarks and a high density
($\rho_n^{cr}$)
phase with light quarks of (partially) restored chiral symmetry
(see also \cite{costa,RSP,buballa}).

With this parameterization \cite{buballa}, $\rho_n^l \simeq 0$, and at zero temperature,
 the model may be interpreted as having a hadronic phase - droplets of light $u\,,d$
quarks with a density
$\rho_n^{cr} = 2.25 \rho_0$ surrounded by a non-trivial vacuum - and, above the
critical density, a quark phase with partially restored chiral symmetry.

The right panel in Figure 1 shows the combined effect of the temperature and density 
dependencies of the pressure.

As  the temperature increases the pressure becomes positive but the compressibility is still negative. At $T=56$ MeV and $\rho=1.53\rho_0$, the compressibility has only on zero and we identify this point as the critical endpoint, the connects the first order and second order phase transition regions.


\section{$\eta$ -- $\eta'$ behavior in the medium}  %

\begin{figure}
\begin{center}
  \begin{tabular}{cc}
    \hspace*{-0.5cm}\epsfig{file=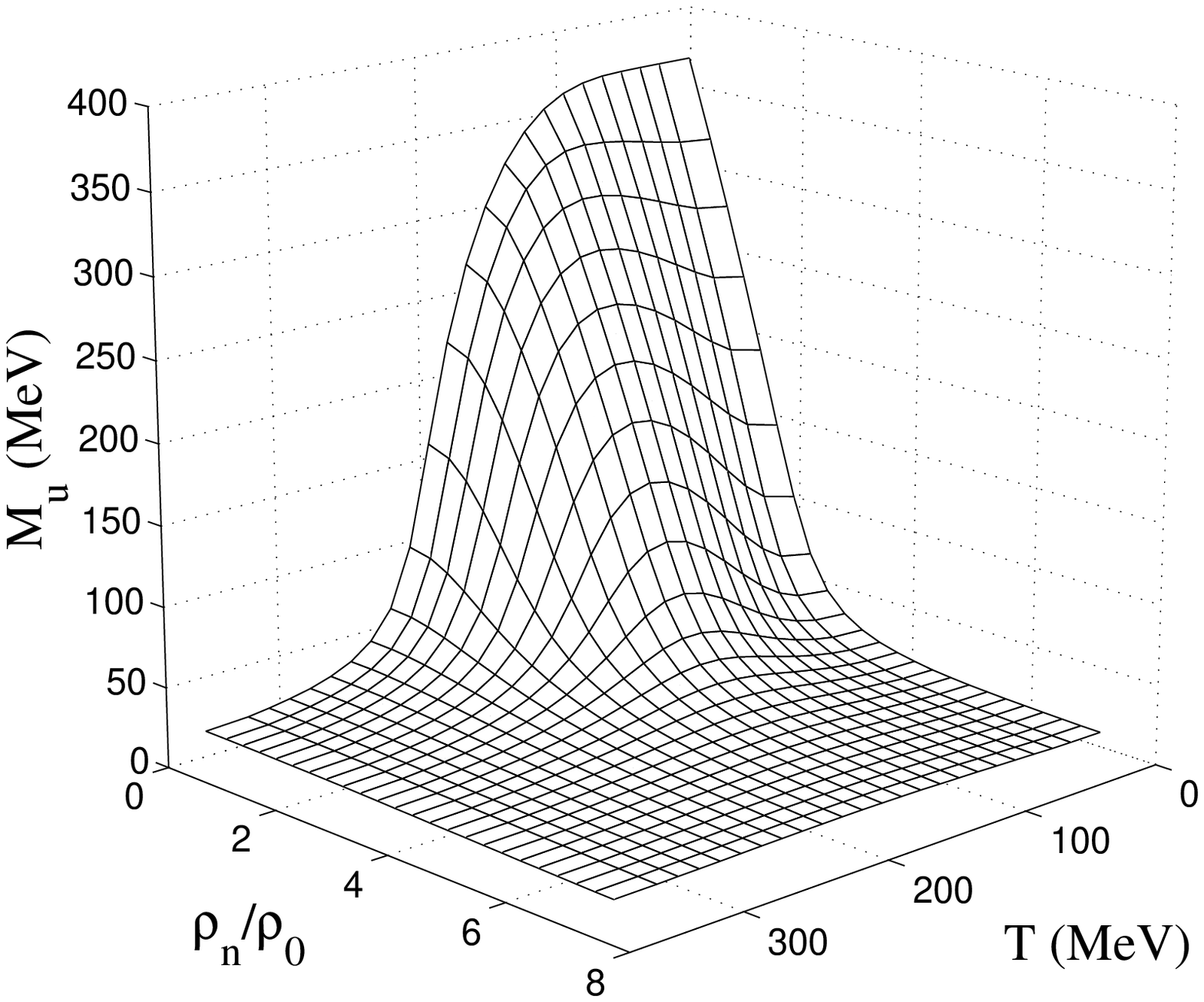,width=9cm,height=7cm} &
    \epsfig{file=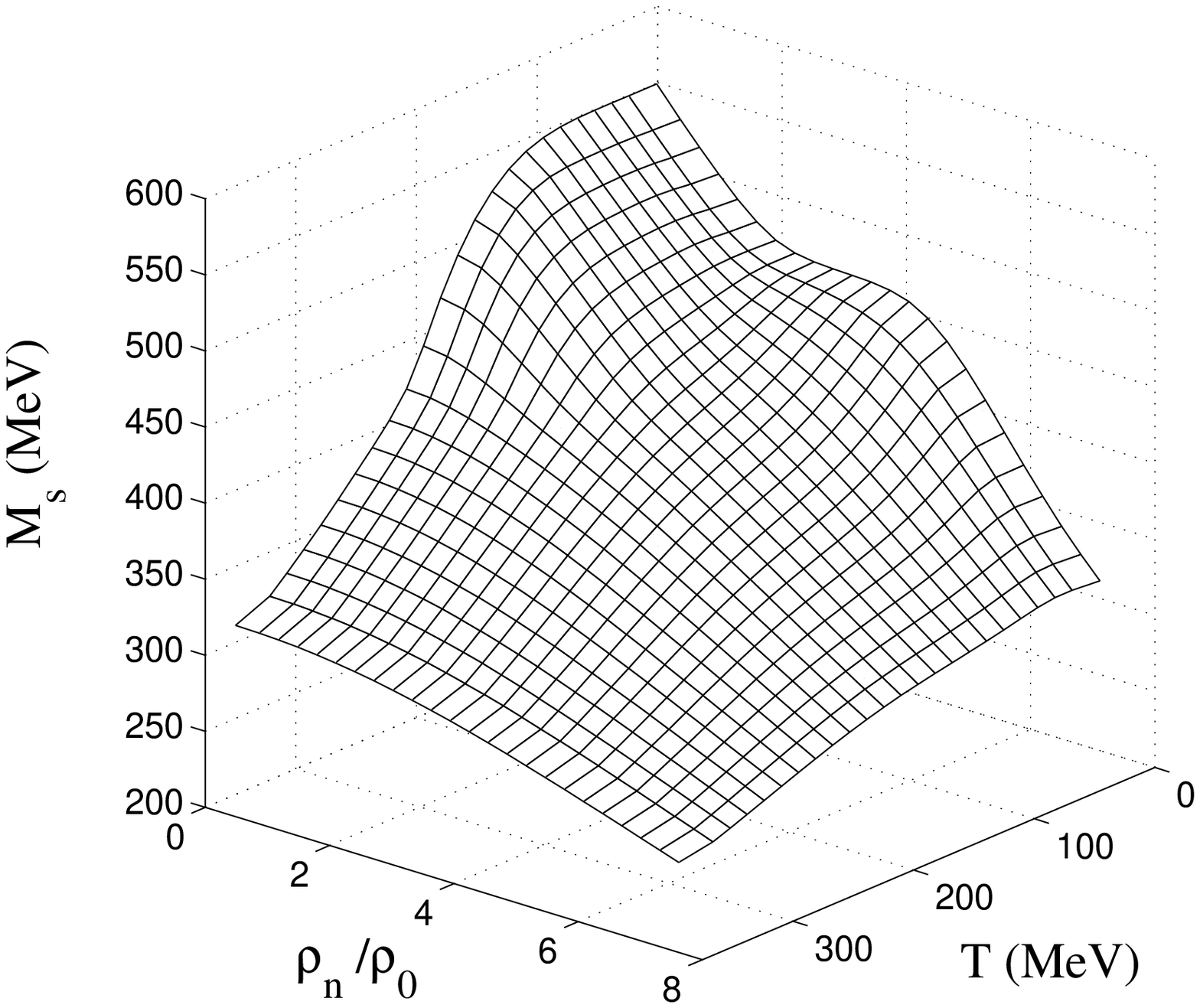,width=9cm,height=7cm} \\
   \end{tabular}
\end{center}

\caption{Masses of the $u$ and $s$ quarks as function of the density and temperature.}
\end{figure}

\hspace*{\parindent}Figure 2 shows the temperature and density dependencies of constituent
quark masses $M_i$ (left panel: $M_u=M_d$, right panel: $M_s$).
As we see at low  temperature masses are almost constant
and then they decrease  with increasing temperature.  
  The same situation can be observed
in the density direction. 

The combined effect of density and temperature shows, as it was expected, that chiral symmetry is partially restored for the light quarks and is badly restored in the strange sector. There is a clear separation between the regions of restored and  chiral symmetry in the diagram of the left panel of Figure 2, but the same does not happen in the right panel.
 
In what concerns the mesonic behavior, before discussing our new findings, we reproduce here the behavior at finite temperature and zero density obtained, for instance in \cite{klev2}.
Figure 3 a) shows the temperature dependence of the $M_\pi$, $M_\eta$ and $M_{\eta'}$ mesons
with vanishing chemical potential. 
For comparison, the curve of $2 M_u$ is also indicated.
One see that at low temperature the $M_\pi$, $M_\eta$
masses are lower than masses of their constituents. In this
case the integrals $I_2^{ii}(P_0, T)$  are real.
The crossing of the $\pi$ and $\eta$ lines with
$2 M_u$  line indicates the respective Mott transition
temperature  $T_M$  for these mesons. For our set of parameters
the mesons become unbound at $T_M \approx 200$ MeV.
One can also see that $T_{M\, \eta} < T_{M\, \pi}$.
Above the Mott temperature we have taken into account the imaginary parts of the integrals
$I_2^{ii}(P_0, T)$ and used a finite width approximation \cite{bla,klev2}.
As we mentioned before the $\eta'$ - meson is unbound.
\begin{figure}
\begin{center}
  \begin{tabular}{cc}
    \epsfig{file=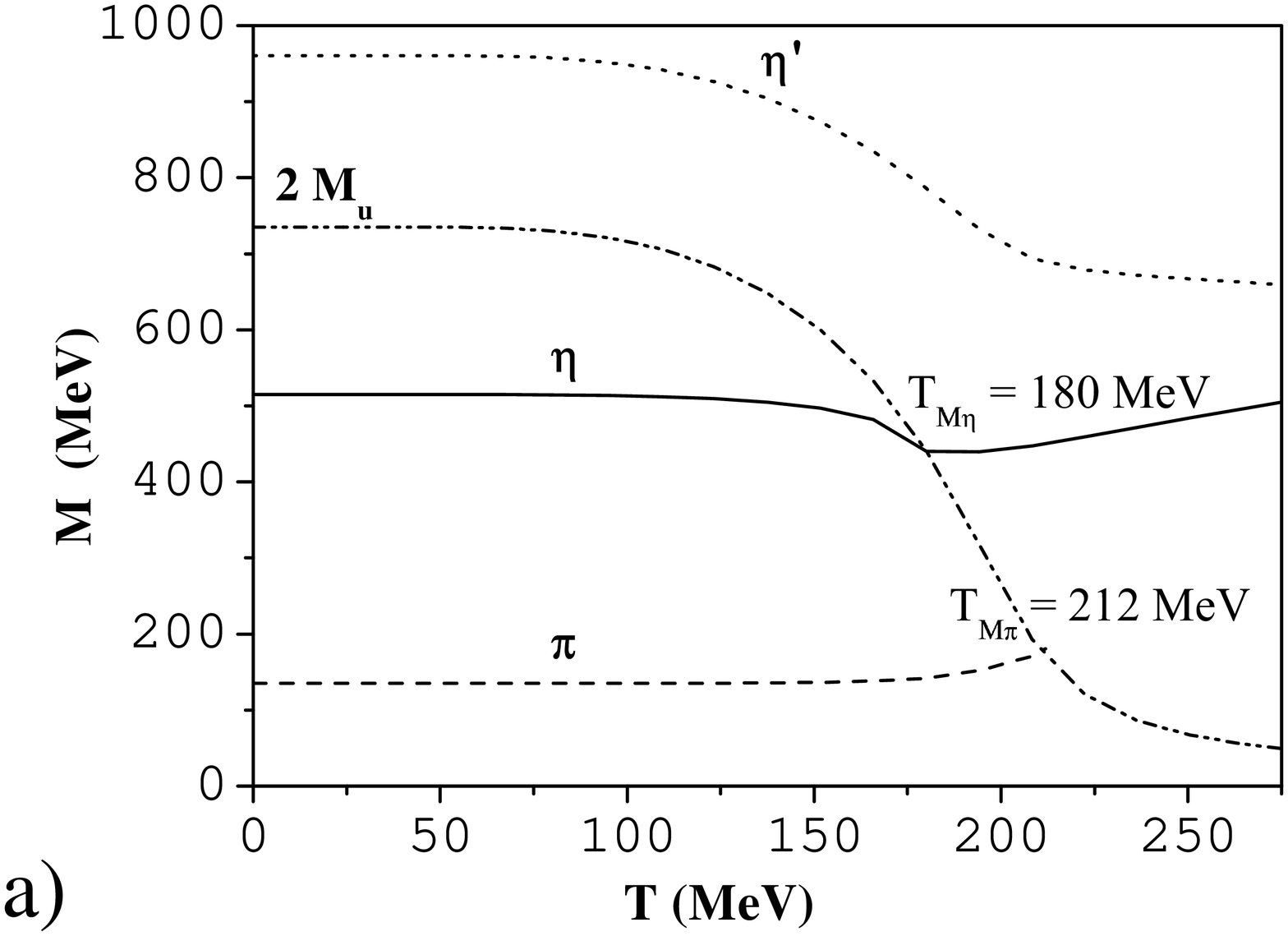,width=9cm,height=6.8cm} &
    \epsfig{file=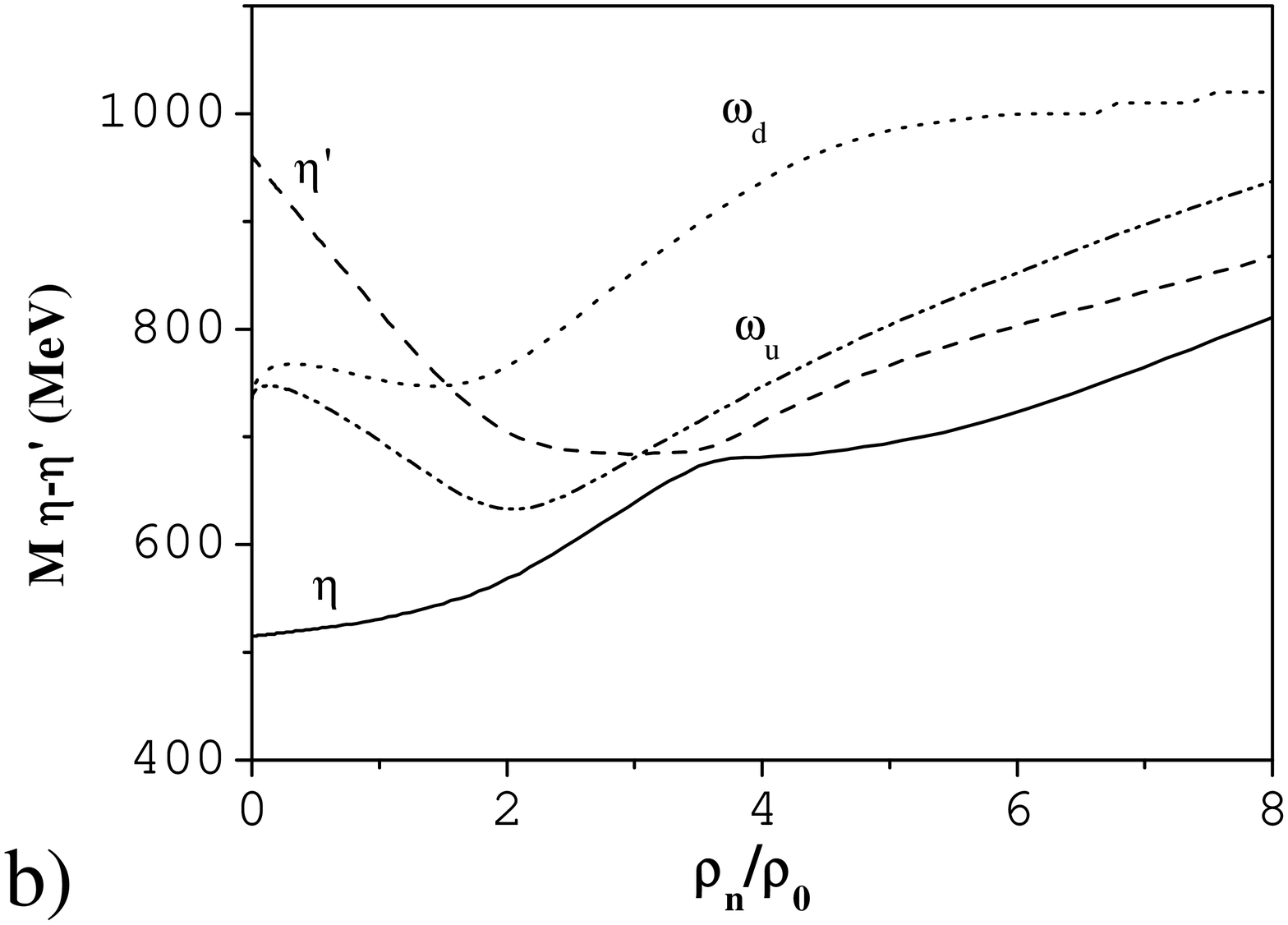,width=9cm,height=6.8cm} \\
   \end{tabular}
\end{center}
\caption{ a) Temperature dependence of the $\pi$ (dashed line),  $\eta$
(solid line) and $\eta'$ (dotted line) masses.
The curve $2 M_u$ shows the temperature dependence of the quark threshold.
Respective Mott temperatures $T_{M \pi}$ and  $T_{M \eta}$ are
indicated; b) Density dependence of  $\eta$
(solid line) and $\eta'$ (dotted line) masses.}
\end{figure}

We start the discussions of the $\eta$ and $\eta'$ in matter
from the properties of dynamical quark masses and chemical potentials.
They  are plotted in Figure 4 as functions of $\rho_n$.

\begin{figure}[ht]
\begin{center}
\epsfig{file=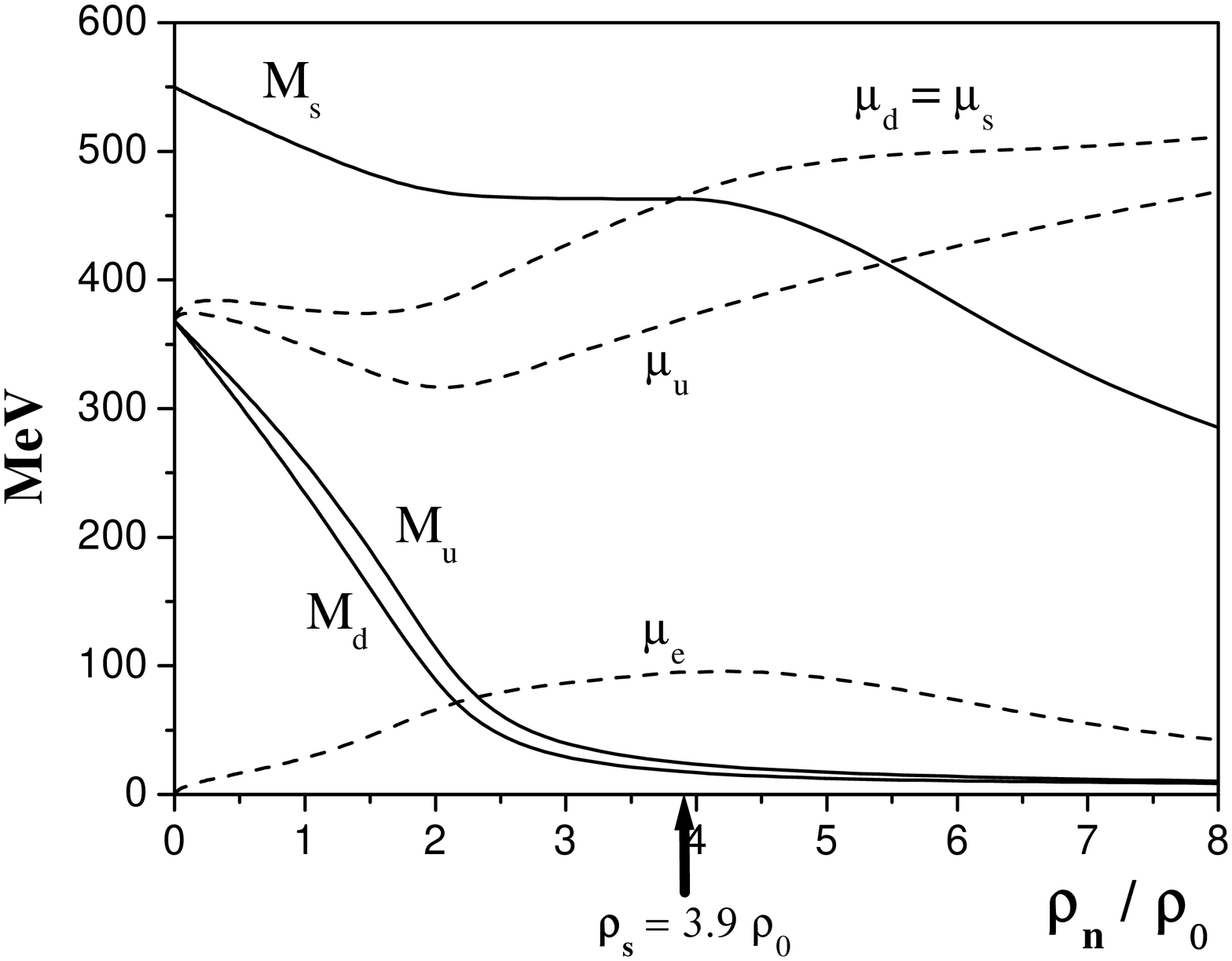, width=12.0cm,height=7cm}
\end{center}
\caption{Constituent quark masses (solid lines) and chemical potentials, $\mu_s=\mu_u\,, \mu_e$ (dashed lines) at
$T = 0$ MeV.}
\end{figure}

An important point to be noticed is that, due to the 't Hooft contribution
in the gap equations, the mass of the strange quark decreases smoothly
and becomes lower than the chemical potential  at densities above 3.9$\rho_0$,
which we will denote from now on as $\rho'_s$. A more pronounced decrease is then
observed, which is due to the presence of $s$ quarks in this regime. As it will be
discussed below, this fact is related to a change in the behavior of different observables,
as compared to the region of lower densities. At $\rho =\rho'_s$ (Figure 4)
the difference between the  chemical potentials $\mu_d=\mu_s$ has a maximum and,
as the density increase there is a  tendency to the restoration of flavor symmetry \cite{costaruivo}.

Now, we analyze the results  for the masses of $\eta$ and $\eta'$.

Having in mind the constraints (\ref{constr1}) and (\ref{constr2})
we have calculated the density dependence of
$M_\eta$ and $M_{\eta'}$ masses. The results are shown on Figure 3-b.

In order to understand these results, it is useful to study the limits
of the Dirac sea of this mesons.
They can be obtained by looking the limits of the regions  of poles in  the
integrals $I_2^{ii} (P_0)$
(\ref{sint}) and are plotted in Figure 3 (dashed point lines): $\omega_{u}= 2\mu_{u}$,
$\omega_{d}= 2\mu_{d}$ and $\omega_{s}= 2\mu_{s}$. (At finite temperature,
we will have $\omega_{u}= 2M_u$,
$\omega_{d}= 2M_d$ and $\omega_{s}= 2M_s$).

As we can see, the $\eta'$ - meson lies above the
quark - antiquark threshold for $\rho_n < 2.5 \rho_0$ and it is a resonant state.
After that density, the $\eta'$ becomes a bound state.

Concerning the masses of $\eta$ and $\eta'$ they exhibit a tendency to became degenerate
but there are no clear indications of restoration of $U_1(A)$ symmetry. In neutron matter with beta equilibrium the systems shows a tendency
to the restoration of  flavor symmetry, which is related with the
presence of strange quarks in the
medium that occurs at about $\rho\sim 3.8 \rho_0$.

Concerning the behavior in a hot and dense medium, a feature to be noticed is that  the $\eta$ meson, that has a Goldstone boson like  nature, shows more clearly the difference between the chiral symmetric and asymmetric phase, as shown in Figure 5.  In conclusion, in the framework of our model, if the behavior of $\eta'$ at high temperatures  and zero density suggests  a possible restoration of the $U_A(1)$ symmetry, the same does not happen when we explore the high density region of a dense or hot and dense medium.

\begin{center}
{\large Acknowledgement:}
\end{center}
Work suported by grant SFRH/BD/3296/2000 (P. Costa), Centro de 
F\'{\i}sica Te\'orica, FCT and GTAE (Yu. L. Kalinovsky).

\newpage
\begin{figure}[ht]
\begin{center}
  \begin{tabular}{cc}
    \hspace*{-0.5cm}\epsfig{file=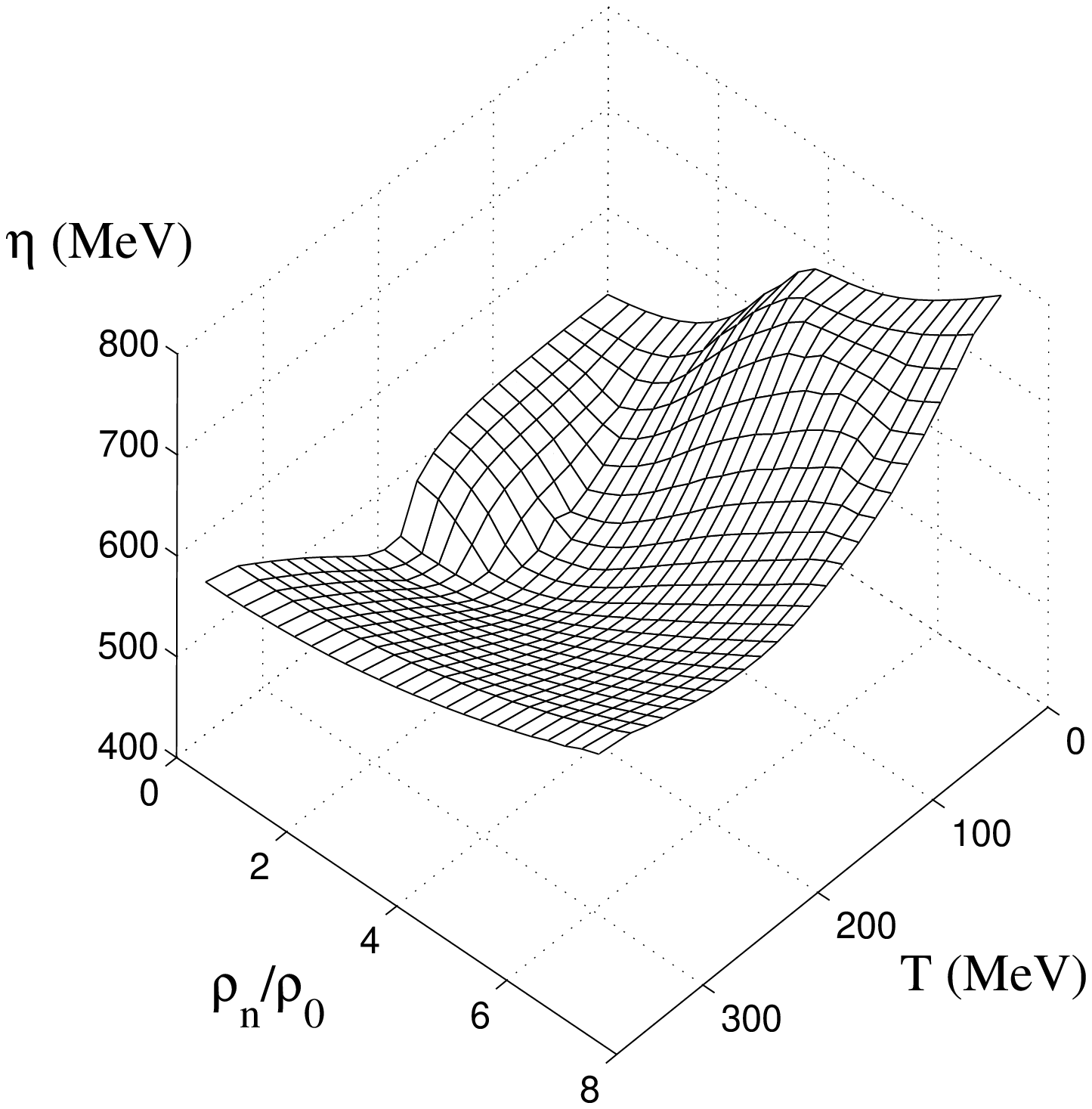,width=9cm,height=8cm} &
    \epsfig{file=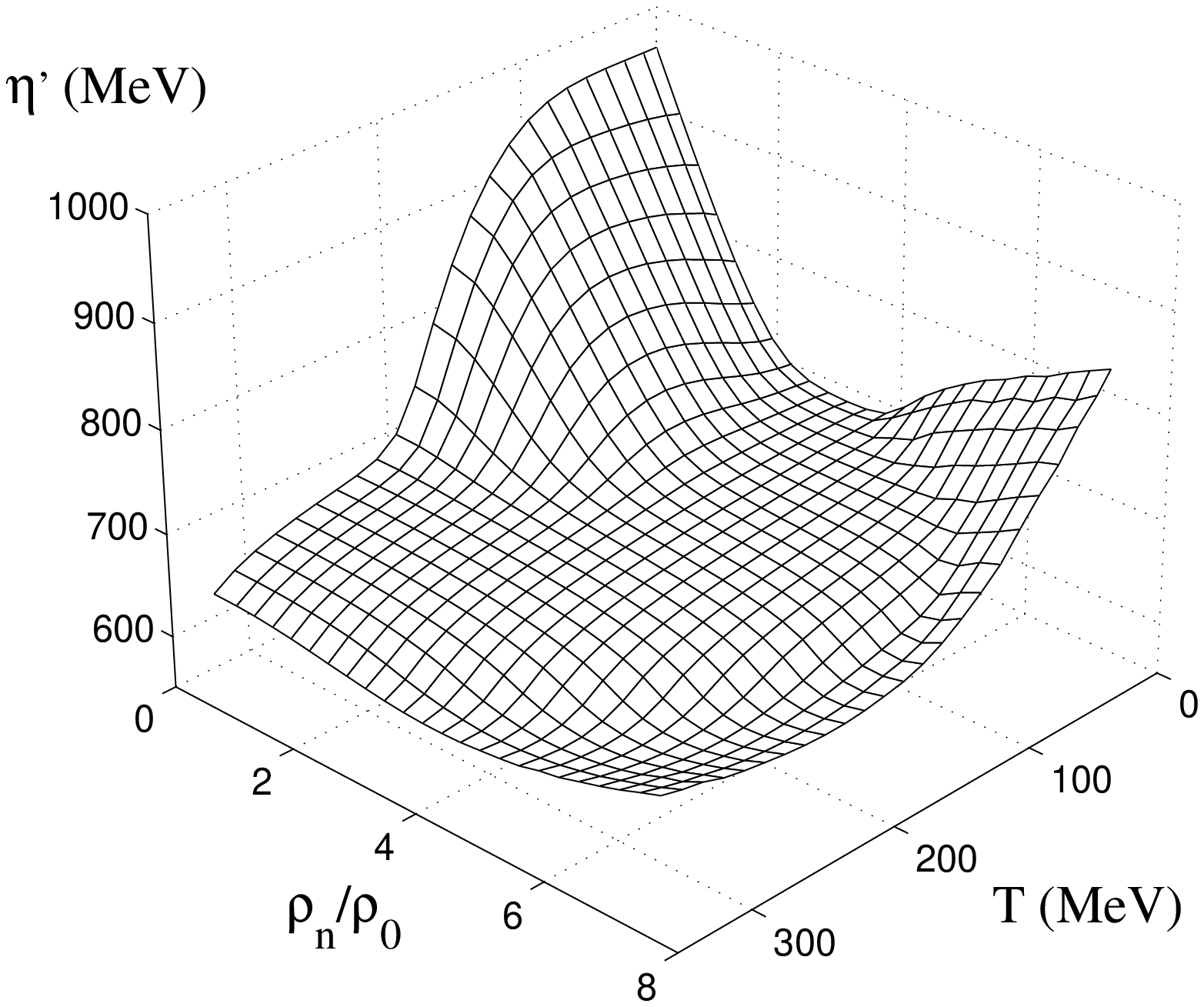,width=9cm,height=8cm} \\
  \end{tabular}
\end{center}
\caption{Left panel: $\eta$ mass as function of the density and temperature;
Right panel: $\eta'$ mass as function of the density and temperature.}
\end{figure}


\begin{thebibliography}{}

\bibitem{kanaya}
{F. Karsch  and E. Laermann}, {Phys. Rev.} {D 50} (1994) {6954}; \\
{K. Kanaya}, {Prog. Theor. Phys. Sup.} {129} (1997) {197}.

\bibitem{rhic} {C. Roland}, (PHOBOS Collaboration), {Nucl. Phys.} {A 698} (2002) {54}.

\bibitem{cern} {C. Louren\c co}, {Nucl. Phys.} {A 698} (2002) {13}.

\bibitem{Fodor} {Z. Fodor and S. D. Katz}, {Phys. Lett.} {B 534} (2002) 87, hep-lat/0104001; hep-lat/0204029.

\bibitem{shuryak} {E. Shuryak}, {Comm. Nucl. Part. Phys.} {21} (1994) 235.

\bibitem{kapusta}  {J. Kapusta, D. Kharzeev and L. McLerran},  {Phys. Rev.} {D 53} (1996) 5028.

\bibitem{njl}
 Y. Nambu and G. Jona-Lasinio, Phys. Rev. 122 (1961) 345; Phys. Rev. 124 (1961) 246; \\ 
M. K. Volkov, Ann. Phys. 157 (1984) 282.

\bibitem{sokol} G. A. Sokol, L. N. Pavlyuchenko, nucl-ex/0111020 (2001).

\bibitem{ohta1}  
K. Ohnishi, K. Fukushima, K. Ohta,    {Phys. Rev.} {C 63} (2001) 045203; {Phys. Lett.} {B 514} (2001) 200.

\bibitem{njlT}
 U. Vogl and W. Weise, Prog. Part. Nucl. Phys. 27 (1991) 195; \\
 S. P. Klevansky, Rev. Mod. Phys. 64 (1992) 649; \\
 T. Hatsuda and T. Kunihiro, Phys. Reports 247 (1994) 221; \\
 P. Rehberg, S. P. Klevansky, Ann. Phys. 252 (1996) 422.

\bibitem{bla}    D. Blaschke et al., Nucl. Phys. A 592 (1995) 561.

\bibitem{maria}  {M. C. Ruivo, C. A. de Sousa, B. Hiller and A. H. Blin}, {Nucl. Phys.} {A 575} (1994) 460.

\bibitem{RuivoSousa} {M. C. Ruivo and C. A. Sousa}, {Phys. Lett.} {B 385} (1996) {39}. 

\bibitem{SousaRuivo}  C. A. Sousa  and M. C. Ruivo, { Nucl. Phys.} {A 625} (1997) 713.  W. Broniowski and B. Hiller, { Nucl. Phys.}  {A 643} (1998) 161.

\bibitem{costa}  M.C. Ruivo, P. Costa  and C. A. Sousa,
   in {\it Quark Matter in Astro and Particle Physics},
   Eds. D. Blaschke, Rostock University, 2001, p. 218, hep-ph/0109234.

\bibitem{RSP} {M. C. Ruivo, C. A. Sousa  and C. Provid\^{e}ncia}, {Nucl. Phys.} {A 651} (1999) {59}.

\bibitem{witten} E. Witten, Nucl. Phys. B 156 (1979)  269.

\bibitem{veneziano} G. Veneziano, Nucl. Phys. B 159 (1979) 213.

\bibitem{Hiller}  W. Broniowski and B. Hiller, { Nucl. Phys.}  {A 643} (1998) 161.

\bibitem{Ruivo} {M. C. Ruivo}, {Hadron Physics - Effective Theories of Low Energy QCD},
{ed. A. Blin et al}, {AIP, New York}, {2000}, {p. 237}.

\bibitem{costaruivo} P. Costa and M. C. Ruivo, Europhysics Letters 60(3) (2002) 356.

\bibitem{buballa}  M. Buballa and M. Oertel, {Nucl. Phys. } {A 642} (1998) 39c;
                  {Phys. Lett. } {B 457} (1999) 261.

\bibitem{klev2}  P. Rehberg, S. P. Klevansky and J. H\"{u}fner, Phys. Rev. C 53 (1996) 410.

\end{thebibliography}
\end{document}